\documentclass[runningheads]{llncs}

\usepackage{amsmath}

\usepackage{graphicx}
\usepackage{listings}
\usepackage{booktabs}
\usepackage{subcaption}
\usepackage{algorithm}
\usepackage{algpseudocode}

\begin{document}
\title{Graph-based Interpretation of Normal Logic Programs}
%
%
\author{Fang Li \and Elmer Salazar \and
Gopal Gupta }
\authorrunning{F. Li, E. Salazar, G. Gupta}
%
\institute{The University of Texas at Dallas, Richardson TX 75080, USA \\
\email{\{fang.li, elmer.salazar, gupta\}@utdallas.edu}}
\maketitle              
\begin{abstract}
In this paper we present a dependency graph-based method for computing the various semantics of normal logic programs. Our method employs \textit{conjunction nodes} to unambiguously represent the dependency graph of normal logic programs. The dependency graph can be transformed suitably in a semantics preserving manner and re-translated into a 
equivalent normal logic program. This transformed normal logic program can be augmented with a few rules written in answer set programming (ASP), and the CLINGO system used to compute its answer sets. Depending on how these additional rules are coded in ASP, one can compute models of the original normal logic program under the stable model semantics, the well-founded semantics, or the co-stable model semantics. In each case, justification for each atom in the model is also generated. We report on the implementation of our method as well as its performance evaluation.
\keywords{Normal Logic Programs, Well-founded Semantics, Stable Model Semantics, Co-stable Model Semantics, Interpretation}

\end{abstract}

\section{Introduction}
Logic programming \cite{abiteboul1995foundations} has been applied to many areas such as fault diagnosis, databases, planning, natural language processing, knowledge representation and reasoning. During decades of exploration, researchers have developed various semantics for solving different reasoning tasks. Among those semantics, the stable model semantics based answer set programming (ASP) \cite{MT5} paradigm is popular for knowledge representation and reasoning as well as for solving combinatorial problems. Though computing ASP programs is considered to be NP-hard, there are a lot of ASP solvers (e.g., CLINGO \cite{gebser2014clingo}, DLV \cite{dlv}, s(CASP) \cite{arias2018constraint}) that can compute stable models of an ASP program efficiently. Meanwhile, there are also many approaches to solve programs under the well-founded semantics, such as XSB \cite{rao1997xsb} and XOLDTNF \cite{chen1993goal}. For the co-stable model semantics \cite{elmerthesis,gupta2011infinite}, there is no specific solving systems designed yet.

In this paper we are interested in computing models of normal logic programs under various semantics: stable model, co-stable model and well-founded semantics, in particular. At the same time, we want to provide justification for inclusion of a given atom in the model. We show that our graph-based approach is able to accomplish both these goals.

At present, to compute the semantics of a normal logic program under different semantics, we have to encode the program in different formats and use different systems to compute the model(s) under various semantics. For finding justification, one may have to use yet another system. Lack of a single system that can achieve all the aforementioned tasks is the main motivation of our research. Graph-based approaches have other additional benefits, such as partial models can be computed which is needed in some applications \cite{li2021discasp}. Elsewhere \cite{li2021graph} the authors introduced a dependency graph-based approach to represent ASP programs and compute their models. The basic idea is to use dependency graph to represent rules and goals of an answer set program, and finding models by reasoning over the graph. This graph-based representation is the basis for our work in this paper in interpreting a normal logic program under different semantics.

Our novel graph-based method for computing the various semantics of a normal logic program uses conjunction nodes to represent the dependency graph of the program. The dependency graph can be transformed in a semantics preserving manner and re-translated into an equivalent normal logic program. This transformed normal logic program can be augmented with a small number of rules written in ASP, and then a traditional SAT solver-based system used to compute its answer sets. Depending on how these additional rules are coded in ASP, one can compute models of the original normal logic program under the stable model semantics, the co-stable model semantics as well as the well-founded semantics. In this paper, we use CLINGO \cite{gebser2014clingo} as the ASP solver to compute models of the augmented normal logic program.

The rest of this paper is structured as follows: Section \ref{background} introduces the background knowledge related to the research. Section \ref{interpretations} presents the interpretation methods for three different semantics. Section \ref{example} uses a simple example to illustrate the work flow of the algorithm. Section \ref{conclusion} concludes the paper and introduces the future work with regard to this research.

\section{Background} \label{background}

\subsection{Answer Set Programming} \label{sec:asp}   

Answer Set Programming (ASP) is a declarative paradigm that extends logic programming with negation-as-failure. ASP is a highly expressive paradigm that can elegantly express complex reasoning methods, including those used by humans, such as default reasoning, deductive and abductive reasoning, counterfactual reasoning, constraint satisfaction~\cite{baral,gelfond2014knowledge}.
ASP supports better semantics for negation ({\it negation as failure}) than does standard logic programming and Prolog. An ASP program consists of rules that look like Prolog rules. The semantics of an ASP program {$\Pi$} is given in terms of the answer sets of the program \texttt{ground($\Pi$)}, where \texttt{ground($\Pi$)} is the program obtained from the substitution of elements of the \textit{Herbrand universe} for variables in $\Pi$~\cite{baral}.
Rules in an ASP program are of the form shown as below (Rule \ref{rule1}):
\begin{equation}
\label{rule1}
    p \; :- \; q_1, \; ..., \; q_m, \; \textbf{not} \; r_1, \; ..., \; \textbf{not} \; r_n.
\end{equation} 
\noindent where $m \geq 0$ and $n \geq 0$. Each of \texttt{p} and \texttt{q$_i$} ($\forall i \leq m$) is a literal, and each \texttt{not r$_j$} ($\forall j \leq n$) is a \textit{naf-literal} (\texttt{not} is a logical connective called \textit{negation-as-failure} or \textit{default negation}). The literal \texttt{not r$_j$} is true if proof of {\tt r$_j$} \textit{fails}. Negation as failure allows us to take actions that are predicated on failure of a proof. 
Thus, the rule {\tt r :- not s.} states that {\tt r} can be inferred if we fail to prove {\tt s}. 
Note that in Rule \ref{rule1}, {\tt p} is optional. Such a headless rule is called a constraint, which states that conjunction of {\tt q$_i$}'s and \texttt{not r$_j$}'s should yield \textit{false}. Thus, the constraint {\tt :- u, v.} states that {\tt u} and {\tt v} cannot be both true simultaneously in any model of the program (called an answer set).

The declarative semantics of an Answer Set Program \texttt{P} is given via the Gelfond-Lifschitz transform~\cite{baral,gelfond2014knowledge} in terms of the answer sets of the program \texttt{ground($\Pi$)}. 
More details on ASP can be found elsewhere~\cite{baral,gelfond2014knowledge}. 

\subsection{Dependency Graph}\label{sec:dg}

A dependency graph \cite{linke2005suitable} uses nodes and directed edges to represent dependency relationships of an ASP rule. 

\begin{definition}
The dependency graph of a program is defined on its literals s.t. there is a positive (resp. negative) edge from $p$ to $q$ if $p$ appears positively (resp. negatively) in the body of a rule with head $q$.
\label{def1}
\end{definition}

\begin{figure}[tb]
\centering
  \centering
  \includegraphics[scale=0.3]{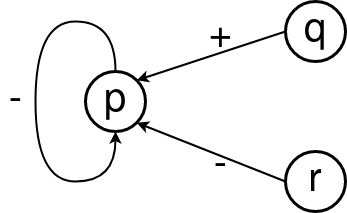}
      \caption{Dep. Graph for Programs 1 \& 2}
    \label{fig:fig1}

  \centering
    \begin{subfigure}[b]{0.5\linewidth}
        \centering
        \includegraphics[scale=0.3]{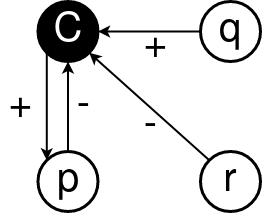}
        \caption{CNR for Program 1}
    \end{subfigure}\hfill
    \begin{subfigure}[b]{0.5\linewidth}
        \centering
        \includegraphics[scale=0.3]{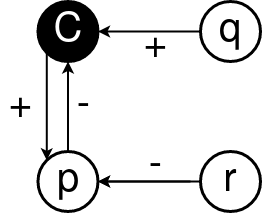}
        \caption{CNR for Program 2}
    \end{subfigure}
    \caption{CNRs for Program 1 \& 2}
    \label{fig:fig2}
\end{figure}

Conventional dependency graphs are not able to represent ASP programs uniquely. This is due to the inability of dependency graphs to distinguish between non-determinism (multiple rules defining a proposition) and conjunctions (multiple conjunctive sub-goals in the body of a rule) in logic programs. For example, the following two programs have identical dependency graphs (Figure \ref{fig:fig1}).

\begin{lstlisting}[language=prolog,basicstyle=\small]
    %% program 1                 %% program 2
    p :- q, not r, not p.        p :- q, not p. p :- not r.
\end{lstlisting}

To make conjunctive relationships representable by dependency graphs, we first transform it slightly to come up with a novel representation method. This new representation method, called conjunction node representation (CNR) graph, uses an artificial node to represent conjunction of sub-goals in the body of a rule. This conjunctive node has a directed edge that points to the rule head (Fig. \ref{fig:fig2}).

The conjunction node, which is colored black, refers to the conjunctive relation between the in-coming edges from nodes representing subogals in the body of a rule. Note that a CNR graph is not a conventional dependency graph.

\subsection{Converting CNR Graph to Dependency Graph} \label{sec:cnrtodg}
Since CNR graph does not follow the dependency graph convention, we need to convert it to a proper dependency graph in order to perform dependency graph-based reasoning. We use a simple technique to convert a CNR graph to an equivalent conventional dependency graph. We negate all in-edges and out-edges of the conjunction node. This process essentially converts a conjunction into a disjunction. Once we do that we can treat the conjunction node as a normal node in a dependency graph. As an example, Figure \ref{fig:fig3} shows the CNR graph to dependency graph transformation for program {\tt p :- q, not r.} This transformation is a simple application of De Morgan's law. The rule in this program represents {\tt p :- C. and C :- q, not r.} The transformation produces the equivalent rules {\tt p :- not C., C :- not q. and C :- r.}


\noindent Since conjunction nodes are just helper nodes which allow us to perform dependency graph reasoning, we don't report them in the final answer set.

\subsection{Constraint Representation} \label{sec:constraintnode}
ASP also allows for special types of rules called constraints. There are two ways to encode constraints: (i) headed constraint where negated head is called directly or indirectly in the body (e.g., Program 3), and (ii) headless constraints (e.g., Program 4). \\
\begin{lstlisting}[language=prolog,basicstyle=\small]
    %% program 3                       %% program 4
    p :- not q, not r, not p.          :- not q, not r.
\end{lstlisting}

\begin{figure}[tb]
\centering
  \centering
  \includegraphics[width=.4\linewidth]{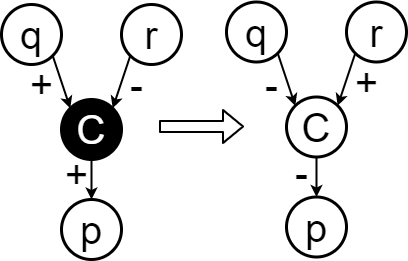}
  \caption{CNR-DG Transformation}
  \label{fig:fig3}
  \centering
    \begin{subfigure}[b]{0.5\linewidth}
        \centering
        \includegraphics[scale=0.3]{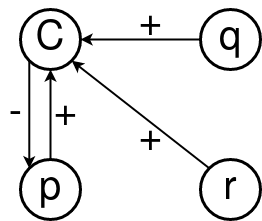}
        \caption{Program 3}
        \label{fig:fig4a}
    \end{subfigure}\hfill
    \begin{subfigure}[b]{0.5\linewidth}
        \centering
        \includegraphics[scale=0.3]{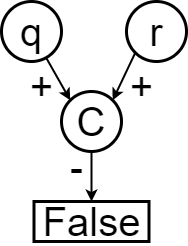}
        \caption{Program 4}
        \label{fig:fig4b}
    \end{subfigure}
    \caption{Constraint DG}
    \label{fig:fig4}
\end{figure}

Our algorithm models these constraint types separately. For the former one, we just need to apply the CNR-DG transformation directly. Note that the head node connects to the conjunction node both with an in-coming edge and an out-going edge (Figure \ref{fig:fig4a}). For the headless constraint, we create a head node with truth value as \textit{False}. The reason why we don't treat a headless constraint the same way as a headed constraint is because in the latter case, if head node ({\tt p} in Program 3) is provable through another rule, then the headed constraint is inapplicable. Therefore, we cannot simply assign a \textit{False} value to its head.

\subsection{Graph-based Transformation of an ASP Program} \label{sec:dg-conversion-asp}
Based on the representation method defined above, we are able to convert an ASP program into the graph format, which uses \textbf{nodes} for goals and \textbf{edges} for rules. This process can be broken down into three steps: (1) converting the original ASP program to CNR representation. (2) converting the CNR graph to dependency graph. (3) wrapping each goal (or rule) with a "node" (or "edge") predicate. Algorithm \ref{alg} is the pseudo-code for implementing this procedure. In our actual implementation, this procedure is written in the RUST language.

\begin{algorithm}[ht]
\small
	\caption{CNR-ASP Parsing Algorithm}
	\begin{algorithmic}[1]
	    \State g $\leftarrow$ new Graph()
	    \State conj\_cnt $\leftarrow$ 0
        \State rules $\leftarrow$ parse(file)
        \For {r in rules} \Comment{parse input program into graph}
            \State head, tail $\leftarrow$ parseRule(r)
            \If{head.len() == 0}
                head $\leftarrow$ "constraint"
            \EndIf
            \If {tail.len() $\leq$ 1}
                \State g.nodes.add(head, true)
            \ElsIf {tail.len() == 1}
                \State g.edges.add(tail, head, tail[0].sign)
            \Else
                \State conj $\leftarrow$ append("conj", ++conj\_cnt.toString())
                \For{t in tail}
                    \State g.edges.add(t, conj, t.sign.negate())
                \EndFor
                \State g.edges.add(conj, head, negative)
            \EndIf
        \EndFor
        \State res $\leftarrow$ new List()
        \For{n in g.nodes} \Comment{wrap goals with nodes}
            \State res.append("node(" + n + ")")
        \EndFor
        \For{e in g.edges} \Comment{wrap rules with edges}
            \State res.append("edge(" + e.from + "," + e.to + "," + e.sign + ")")
        \EndFor
        \State \Return {res}
	\end{algorithmic}
	\label{alg}
\end{algorithm}

Let's take Program 5 as an example. 

\begin{lstlisting}[language=prolog,basicstyle=\small]
    %% program 5
    p :- not q, r. q :- not p.
\end{lstlisting}

\noindent 
Step 1, we need to use CNR representation to solve the conjunctive goals in the body of the first rule:

\begin{lstlisting}[language=prolog,basicstyle=\small]
    negative_edge(q, conjunct_1).
    positive_edge(r, conjunct_1).
    positive_edge(conjunct_1, p).
\end{lstlisting}

\noindent 
The second rule will be directly represented as

\begin{lstlisting}[language=prolog,basicstyle=\small]
    negative_edge(p, q).
\end{lstlisting}

\medskip\noindent 
Step 2, negating the signs of each edge (with conjunct\_1) to get a dependency graph: 

\begin{lstlisting}[language=prolog,basicstyle=\small]
    positive_edge(q, conjunct_1).
    negative_edge(r, conjunct_1).
    negative_edge(conjunct_1, p).
    negative_edge(p, q).
\end{lstlisting}

\medskip\noindent
Step 3, wrapping goals and rules: Here we should parameterize the edge signs for the reasoning purpose. Therefore, the final dependency graph representation will be:

\begin{lstlisting}[language=prolog,basicstyle=\small]
    node(p). node(q). node(r). node(conjunct_1).
    edge(q, conjunct_1, positive).
    edge(r, conjunct_1, negative).
    edge(conjunct_1, p, negative).
    edge(p, q, positive).
\end{lstlisting}

\subsection{Effective Edge} \label{bg:effective_edge}
An effective edge in a dependency graph refers to any edge that propagates \textit{True} value to the node it is incident on. There are two types of \textit{effective} edges: (i) positive edge emanating from a \textit{True} node; (ii) negative edge emanating from a \textit{False} node. An effective edge only points to a \textit{True} node.

\subsection{Loops in Programs}
In an ASP program, loops among literals may exist. There are three kinds of loops that can be found in the program: even loops, odd loops, and positive loops. Even loops and odd loops refer to loops that have an even or odd number of negative edges in the corresponding dependency graph. Positive loops are loops with no negative edge. 

\section{Graph-based Normal Logic Program Interpretations} \label{interpretations}
To compute different semantics of a normal logic program, we transform it into a graph-based representation (discussed in Section \ref{sec:dg-conversion-asp}), then augment the program with the corresponding interpreter (consisting of ASP rules). The resulting augmented ASP program can be solved by any ASP solver system (e.g., CLINGO). In this section, we introduce three interpreters which turn CLINGO into solvers for different semantics.

Note that information regarding dependence between various literals in the program is made explicit in the graph representation. This information is maintained in the ASP program that results from the transformation and can be made explicit in the answer sets of this transformed program. This explicit information on dependence can then be used to provide justification automatically.

It should be noted that this graph-based transformation approach can be quite useful for the analysis of normal logic programs. In general, some interesting property of the program can be computed and the dependency graph annotated. The annotated graph can then be transformed into an answer set program with the annotation intact. Models of the transformed program can then be computed using traditional solvers. The annotations in the model can then be used for computing the program property of interest. Finding causal justification of literals in the model is but one example of this general technique.

\subsection{CNR Graph Transformation} \label{flaw}
It is worth mentioning that our CNR based dependency graph transformation of normal logic programs has an inherent ``flaw". However, this flaw actually works to our advantage. As introduced in previous sections, the CNR approach makes use of De Morgan' Law by inserting an extra node with negated in/out edges to convert conjunctive relationships into disjunctions. From the perspective of logical meaning, double-negation is equivalent to a positive claim, but under the stable model semantics, it may lead to differences. 

For example, program {\small {\tt p :- q, r. q :- p. r.}} has a positive loop between $p$ and $q$. According to the stable model semantics, this program will only have an empty model, in which both $p$ and $q$ are false. But after adding the double-negation based conjunctive node $conj\_1$ to represent {\tt p :- q, r.}, the new graph will form an even loop between $p$, $conj\_1$ and $q$. Since we have $r$ as a fact, under this case, we will get two models \{p/True, q/True\} and \{p/False, q/False\}, which violate the stable model semantics! But at the same time, this follows the co-stable model semantics (introduced later).

This ``flaw" in our graph transformation actually allows us to compute models under the the co-stable model semantics. Therefore, all of our interpreters that are introduced in this paper will be based on the co-stable model semantics interpretation. For interpreting the stable model semantics, we just need to bypass the double-negation related to conjunctive nodes in order to falsify positive loops. For the well-founded semantics, we will have to change the values of each node from binary to ternary (\textit{True, False}, and \textit{Unknown}).

\subsection{The Co-stable Model Semantics Interpretation}
Let us next introduce the co-stable model semantics interpretation, because it is the basis of the other two interpreters. 

Co-stable model semantics  \cite{elmerthesis,gupta2011infinite} is similar to the stable model semantics, except in the last step of the Gelfond-Lifschitz transform, where, instead of finding the least fixpoint of the residual program, we compute its greatest fixpoint. 
Thus, the co-stable model semantics is similar to stable model semantics, the only difference is the way it handles positive loops. In the co-stable model semantics, a positive loop generates two models, consisting of an empty model and an all-true model. For example, in program {\tt p :- q. q :- p.}, co-stable model semantics will return \{p/True, q/True\} and \{p/False, q/False\}.

Co-stable model semantics is useful in many situations where we have cyclical dependency between positive goals. For instance, consider the following two rules: (i) Jack will drink wine if Jill drinks wine, and (ii) Jill will drink wine if Jack drinks wine. In this case, co-stable model semantics will produce the two expected answer sets: one in which both Jack and Jill drink wine, and another in which none of the two drink wine (stable model semantics will produce only a single model, namely, none of the two will drink wine).

The following ASP rules interpret the co-stable model semantics:

\begin{lstlisting}[language=prolog,basicstyle=\small, numbers=left, columns= flexible, keywords={not}]
effective_edge(X,Y) :- edge(X,Y,positive), not false(X).
effective_edge(X,Y) :- edge(X,Y,negative), false(X).
    
true(X) :- fact(X).
true(X) :- node(X), can_pos(X), not false(X).

false(X) :- node(X), not can_pos(X), not true(X).

can_pos(X) :- edge(Y,X,_), effective_edge(Y,X).

:- true(constraint).

#show true/1.
#show false/1.

\end{lstlisting}

This interpreter only has 6 rules, which directly reflects the CNR-dependency graph representation of a normal logic program. As mentioned in section \ref{flaw}, the CNR-dependency graph representation changes a positive loop to an even loop. Therefore, in the interpreter, we only need to define the rules for realizing effective edges (i.e., assign \textit{True} value to the tail of each effective edge). Lines 1-2 define the rules for effective edges: (i) when the head of an edge is not \textit{False} and the edge sign is positive, then the edge is effective; (ii) when the head of an edge is \textit{False} and the edge sign is negative, then the edge is effective. Line 4-5 defines the situations that make a node to be labeled \textit{True}: (i) the node is a fact in the original program; (ii) the node can be positive, and cannot be proved as \textit{False}. Line 7 defines when a node is to be labeled \textit{False}, which is in a manner opposite of the previous rule. Finally, line 9 defines the situation in which a node can be assigned \textit{True}, that is 
the node being the tail of an effective edge. Line 11 ensures that the constraint node is always labeled \textit{False}. Line 13-14 set parameters for CLINGO to only print the \textit{True} or \textit{False} nodes.

\subsection{The Stable Model Semantics Interpretation}
The stable model semantics is the basis of answer set programming. In this semantics, even loops generate multiple worlds, while odd loops kill worlds. For example, in program {\small {\tt p :- not q. q :- not p.}}, $p$ and $q$ form an even loop, which generates two mutually exclusive worlds: \{p/True, q/False\} and \{q/True, p/False\}. For program {\small {\tt p :- not q. q :- not r. r :- not p.}}, nodes $p$, $q$ and $r$ form an odd loop, which makes the program unsatisfiable. A positive loop only generates an empty model.

The following ASP rules interpret the stable model semantics:

\begin{lstlisting}[language=prolog,basicstyle=\small, numbers=left, columns= flexible, keywords={not}]
effective_edge(X,Y,positive) :- edge(X,Y,positive), not false(X).
effective_edge(X,Y,negative) :- edge(X,Y,negative), false(X).

true(X) :- fact(X).
true(X) :- node(X), can_pos(X), not false(X).

false(X) :- node(X), not can_pos(X), not true(X).

can_pos(X) :- edge(Y,X,_), effective_edge(Y,X, Sign).

negate(positive, negative).
negate(negative, positive).

update(positive, negative, negative).
update(negative, positive, negative).
update(positive, positive, positive).
update(negative, negative, negative).

depends(X,Y,Sign) :- effective_edge(Y,X,Sign).
depends(X,Y,Sign) :- not conjunct(Z), effective_edge(Z,X,positive),
    depends(Z,Y,Sign).
depends(X,Y,Sign) :- conjunct(Z), effective_edge(Z,X,negative), edge(Z2,Z,S2), 
    negate(S2,S3),depends(Z2,Y,S4), update(S3,S4,Sign).

:- true(N), depends(N,N,positive).
:- true(constraint).

#show true/1.
#show false/1.
\end{lstlisting}

Based on the co-stable model interpreter, we added some rules to determine whether an even loop is native or caused by the CNR transformation. In the later case, there should be a conjunction node involved. Line 19-25 define rules to extinguish real positive loops from even loops that are caused by the CNR transformation. By proving \textit{depends/3}, we will know whether a node depends on itself and the sign of the dependency. If a node depends on itself and the sign of the dependency is positive, it means that the node is in a positive loop.

Basically, this stable model semantics interpreter extended the co-stable model semantics interpreter, adding the positive loop detection mechanism to prevent the extra model they generate.

\subsection{The Well-founded Semantics Interpretation}
The well-founded semantics can be viewed as a three-valued version of the stable model semantics. Instead of only assigning propositions \textit{True} or \textit{False}, it also allows for a value representing ignorance (\textit{Unknown}). Any atom in an even loop or an odd loop will be considered as \textit{Unknown}.
The following ASP rules interpret the well-founded semantics:

\begin{lstlisting}[language=prolog,basicstyle=\small, numbers=left, columns= flexible, keywords={not}]
effective_edge(X,Y,positive) :- edge(X,Y,positive), not not_true(X).
effective_edge(X,Y,negative) :- edge(X,Y,negative), false(X).

not_true(X) :- node(X), not true(X).

true(X) :- fact(X).
true(X) :- node(X), can_pos(X), not false(X), not unknown(X).

false(X) :- node(X), not can_pos(X), not true(X), not unknown(X).

unknown(X) :- node(X), not true(X), not false(X).

can_pos(X) :- edge(Y,X,_), effective_edge(Y,X, Sign).
can_unknown(X) :- node(X), not can_pos(X), edge_type(S),
    edge(Y,X,S), unknown(Y).

negate(positive, negative).
negate(negative, positive).

edge_type(positive).
edge_type(negative).

update(negative, S, negative) :- edge_type(S).
update(positive, S, S) :- edge_type(S).

dependent_edge(X,Y,Sign) :- true(Y), effective_edge(X,Y,Sign).
dependent_edge(X,Y,Sign) :- unknown(X), unknown(Y), can_unknown(Y),
    edge(X,Y,Sign).
dependent_edge(X,Y,Sign) :- false(Y), not can_unknown(Y), edge(X,Y,Sign).

depends(X,Y,Sign) :- not conjunct(Y), dependent_edge(Y,X,Sign).
depends(X,Y,Sign) :- not conjunct(Z), dependent_edge(Z,X,S1), 
    depends(Z,Y,S2), update(S1,S2,Sign).
depends(X,Y,Sign) :- conjunct(Z), dependent_edge(Z,X,negative), 
    dependent_edge(Z2,Z,S1), negate(S1,S2), depends(Z2,Y,S3), update(S3,S2,Sign).

:- edge(N1,N2,S), edge_type(S), false(N2), unknown(N1).
:- edge(N1,N2,S), edge_type(S), unknown(N2), effective_edge(N1,N2,S).
:- node(N), not unknown(N), not conjunct(N), depends(N,N,negative).
:- node(N), not false(N), depends(N,N,positive), not depends(N,N,negative).
:- unknown(N), not can_unknown(N).
:- not unknown(constraint).

#show true/1.
#show false/1.
#show unknown/1.
\end{lstlisting}

Besides assigning \textit{False} to positive loop nodes, the well-founded semantics also gives \textit{Unknown} value to nodes involved in even loops and odd loops. So the interpreter will be built on the stable model semantics interpretation rules. 

The first thing we need to handle is the representation of the \textit{Unknown} value. In line 1, we use {\tt not not\_true(X)} to define effective edges instead of {\tt not false(X) } in the other two interpreters. Then in line 11, \textit{Unknown} is defined by neither \textit{True} nor \textit{False}. By doing this, we can avoid the even loop between {\tt true(X)} and {\tt false(X)}, so that the domain of the value of a node will have three values. Line 26-35 redefine the {\tt depends/3} from the stable model interpreter by adding aspects related to the \textit{Unknown} value.

\section{Examples and Results} \label{example}
In this section we will use an example to illustrate the process of computing models for a simple program under the three different semantics. First,  we transform the program into the dependency graph representation, and then regenerate it into an answer set program in the node/edge format. The transformed program is then augmented with each of the rule-sets corresponding to three different semantics. Finally, the augmented programs is evaluated with CLINGO to obtain the answer sets. Here is a simple example answer set program:

\begin{lstlisting}[language=prolog,basicstyle=\small]
    p :- not q, r.
    q :- not p.
    r :- p.
\end{lstlisting}

After transformation, the program will appear as follows:

\begin{lstlisting}[language=prolog,basicstyle=\small]
    node(p).
    node(conjunct(0)).
    node(q).
    node(r).
    conjunct(conjunct(0)).
    edge(p,q,negative).
    edge(p,r,positive).
    edge(conjunct(0),p,negative).
    edge(q,conjunct(0),positive).
    edge(r,conjunct(0),negative).
\end{lstlisting}

When augmented with the stable model interpretation rules, CLINGO will produce a single model: \{false(p), true(conjunct(0)), true(q), false(r)\}. This is because in the stable model semantics, the positive loop between node $p$ and $r$ cannot make any of them to be \textit{True}. Only $q$ being \textit{True} can make all rules consistent.

When the transformed program is appended with the co-stable model interpretation rules, CLINGO will produce two models for it: \{false(p), true(conjunct(0)), true(q), false(r)\} and \{false(q), true(p), false(conjunct(0)), true(r)\}. That is because the positive loop allows both $p$ and $r$ becoming \textit{True} simultaneously in the second answer set.

When combining with the well-founded semantics interpretation rules, CLINGO returns one model: \{unknown(p), unknown(conjunct(0)), unknown(q), unknown(r)\}. That is because $p$, $q$, and $r$ are all entangled in loops with negation and therefore are unknown.

\section{Performance}

The interpreters for various semantics introduced in this paper are built on top of CLINGO, so the overhead incurred needs to be understood.
To quantify this overhead, we compare the running time of the interpreted program with running time when the program is directly run on CLINGO without transformation. Of course, we can do this exercise only for the stable model semantics, as CLINGO does not have the option for computing the co-stable model semantics or the well-founded semantics. 

\begin{table} [ht]
\centering
\begin{tabular}{l c|c}
Problem  & Original (seconds) & Interpreted (seconds) \\
\hline
Coloring (4 nodes) & 0.001  & 0.007\\
Coloring (10 nodes) & 0.004  & 0.022\\
Ham Cycle (4 nodes, not fully connected) & 0.002  & 0.006\\
Ham Cycle (4 nodes, fully connected) & 0.003  & 0.011\\
N Queens (N = 4)    & 0.004 & 0.009\\
N Queens (N = 8)   & 0.037 & 0.082\\
K Clique (5 nodes)    & 0.001 & 0.002 \\
K Clique (10 nodes)    & 0.001 & 0.005 \\
\hline
\end{tabular}
\caption{Performance Comparison on Classic Problems}
\label{tb1}
\end{table}

From table \ref{tb1} we can see that the interpreted approach is slower than directly computing the model using CLINGO. The overhead is anywhere from two to five times. The upside is that we gain the ability to compute the models for normal logic programs under different semantics, as well as provide justification (discussed next) for a given answer set under any of the three semantics.

\section{Model Justification}
One important feature of the graph-based interpretation method is the convenience of being able to justify why an atom appears or does not appear in a model. Since all the goals and their connecting edges are shown in the graph, their causal relationships are explicitly captured in the dependence graph representation. Therefore, when justifying a model, we just need to validate all of the effective edges (discussed in section \ref{bg:effective_edge}). These effective edges capture the justification for each of the atoms in a model. 
Let us take the example from section \ref{example}. We know that the graph contains 5 edges: 

\begin{lstlisting}[language=prolog, numbers=left,basicstyle=\small]
    edge(p,q,negative).
    edge(p,r,positive).
    edge(conjunct(0),p,negative).
    edge(q,conjunct(0),positive).
    edge(r,conjunct(0),negative).
\end{lstlisting}

\noindent\textbf{Justifying the Stable Models:}
For the stable model semantics case, we only have one model: \{false(p), true(conjunct(0)), true(q), false(r)\}. Plugging these nodes into each edge, we will get three effective edges: \{edge 1, edge 4, edge 5\}. Accordingly, the end points of those effective edges $q$ and $conjunct(0)$ should also be \textit{True}. Since $q$ and $conjunct(0)$ are both in the model, the model is justified. This causal justification can be output in a user-friendly format.

\medskip 
\noindent\textbf{Justifying the Co-stable Models:}
For the co-stable model semantics case, we have two models. The first model is identical to the stable model, therefore, the justification is omitted here.
The other model is \{false(q), true(p), false(conjunct(0)), true(r)\}. After plugging in to the edges relation, we get two effective edges \{edge 2, edge 3\}. Then from the ending points of effective edges, we get \{true(r), true(p)\} which are in the model. Thus, this model is also valid.

\medskip 
\noindent\textbf{Justifying the Well-founded Model}
Under the well-founded semantics, there will always be only one model for a program. In this example, we have \{unknown(p), unknown(conjunct(0)), unknown(q), unknown(r)\}, which means no effective edges could be formed. Therefore, there is no \textit{True} node in this model. Thus, the model is valid.

\section{Related Work}
To the best of our knowledge, the method that is introduced in this paper---which interprets a normal logic program under different semantics using a traditional ASP solver---is novel. Since our method also provides justification for ASP models, we give a brief overview of relevant approaches to justification for ASP.

Off-line and on-line justifications \cite{pontelli_son_elkhatib_2009} provide a graph-based explanation of the truth value (i.e., true, false, or assume) of a literal. The explanation \textit{assume} is used for literals whose truth value is not being requested. Causal Graph Justification \cite{cabalar_fandinno_fink_2014} explains why a literal is contained in an answer set, but not why a negated literal is not contained. LABAS \cite{schulz_toni_2016} explains the truth value of an extended goal with respect to a given answer set. A goal is in the answer set if a derivation of this goal is supported and is not in the answer set if all derivations of this goal are “attacked”. XSB \cite{rao1997xsb} based ErgoAI (https://coherentknowledge.com) generates justification trees for programs with variables, but it only works with the well-founded semantics. s(CASP) \cite{arias2018constraint} can generate justifications for Constraint Answer Set programs, preventing generating excessively many justifications thanks to its ground-free, top-down evaluation strategy and the use of constraints. It can also justify negated literals and global constraints.

Comparing to the above approaches, the justification method we present in this paper has two major advantages: (i) the graph preserves the causal relationships that are defined by the program rules, but doesn't need to maintain a causal chain (as in the Causal Graph Justification approach) to justify a literal in the model. (ii) it is capable of justifying models under different semantics, instead of only being bound to justification for ASP under stable model semantics.

\section{Conclusion and Future Work} \label{conclusion}
In this paper, we presented a novel method to represent normal logic programs as dependency graphs. After augmenting with ASP interpretation rules, these programs can be executed under different semantics by an ASP solver. This research provides a new perspective, that a normal logic program can be transformed into an intermediate format, which, when combined with specific interpreters can be used for computing semantics other than stable model semantics using the extremely efficient solver developed for answer set programming such as CLINGO. Thus, the efficiency of ASP solvers is exploited to develop interpreters for normal logic programs wrt other semantics with very little effort. The graph-based method also allows us to find the justification for the model(s) computed.

In the future, we plan to work on optimizing our dependence graph-based implementation. Since interpretation rules actually define a symbolic graph structure of a semantics, it means that all programs under one semantics will have the same abstract graph structure. Therefore, we can optimize a solver exclusively for each such graph structure. This is akin to specializing (partially evaluating) an interpreter w.r.t. a given program.

\medskip  
\noindent\textbf{Acknowledgement:}
Authors gratefully acknowledge support from NSF grants IIS 1718945, IIS 1910131, IIP 1916206, and from Amazon Corp and US DoD. 

\bibliographystyle{splncs04}
\bibliography{myilp}   

\end{document}